\documentclass[aps,prd,twocolumn,10pt,floatfix,superscriptaddress,nofootinbib,eqsecnum]{revtex4-2}

\usepackage{bm}
\usepackage{amsmath}
\usepackage{graphicx}
\usepackage{microtype}
\usepackage{latexsym}
\usepackage{amsfonts}
\usepackage{mathrsfs}
\usepackage{color}
\usepackage{hyperref}

\hypersetup{
	hidelinks,
}

\usepackage[T1]{fontenc}
\usepackage{lmodern,amssymb,mathtools}
\usepackage{xcolor,booktabs}
\usepackage{bookmark}
\hypersetup{pdftitle={A Bonnor dihole in a swirling universe},pdfsubject={Einstein-Maxwell equilibrium dihole}}
\newcommand{\dd}{\mathrm d}
\newcommand{\eps}{\epsilon}
\newcommand{\Area}{\mathcal A}

\begin{document}

\title{A Bonnor dihole in a swirling universe:\texorpdfstring{\\}{ }equilibrium of oppositely charged extremal black holes}
\author{Hanjun Zou}
\email{zouhanjun24@mails.ucas.ac.cn}
\affiliation{School of Fundamental Physics and Mathematical Sciences, Hangzhou Institute for Advanced Study, UCAS, Hangzhou 310024, China}
\affiliation{Taiji Laboratory for Gravitational Wave Universe (Beijing/Hangzhou), University of Chinese Academy of Sciences, Beijing 100049, China}
\affiliation{Institute of Theoretical Physics, Chinese Academy of Sciences, Beijing 100190, China}

\author{Chenghao Yuan}
\email{yuanchenghao24@mails.ucas.ac.cn}
\affiliation{School of Fundamental Physics and Mathematical Sciences, Hangzhou Institute for Advanced Study, UCAS, Hangzhou 310024, China}
\affiliation{Taiji Laboratory for Gravitational Wave Universe (Beijing/Hangzhou), University of Chinese Academy of Sciences, Beijing 100049, China}
\affiliation{Institute of Theoretical Physics, Chinese Academy of Sciences, Beijing 100190, China}

\author{Ruo-Cen Wu}
\email{wuruocen24@mails.ucas.ac.cn}
\affiliation{School of Fundamental Physics and Mathematical Sciences, Hangzhou Institute for Advanced Study, UCAS, Hangzhou 310024, China}
\affiliation{Taiji Laboratory for Gravitational Wave Universe (Beijing/Hangzhou), University of Chinese Academy of Sciences, Beijing 100049, China}
\affiliation{Institute of Theoretical Physics, Chinese Academy of Sciences, Beijing 100190, China}

\author{Zhoujian Cao}
\email[Corresponding author: ]{zjcao@bnu.edu.cn}
\affiliation{School of Fundamental Physics and Mathematical Sciences, Hangzhou Institute for Advanced Study, UCAS, Hangzhou 310024, China}
\affiliation{School of Physics and Astronomy, Beijing Normal University, Beijing 100875, China}

\begin{abstract}
We present what is, to our knowledge, the first exact stationary
Einstein--Maxwell solution describing a balanced pair of counterrotating
extremal black holes with opposite electric and magnetic charges in a
gravitational swirling background. An axial Ehlers transformation of the
magnetostatic Bonnor dihole generates the horizon spins and the background
swirl. Tuning the swirl strength removes conical singularities along the
rotation axis. The balanced family has two continuous parameters.
Its exterior is free of closed causal curves,
and the metric and Maxwell field extend smoothly across both horizons.
We give the explicit metric and electromagnetic potentials,
compute the horizon charges and thermodynamic quantities, and verify the
Smarr relation and the first law of black-hole thermodynamics along the
balanced family. The complete near-horizon metric and Maxwell field match
those of extremal dyonic Kerr--Newman, and the microscopic entropy computed
through the Kerr/CFT correspondence reproduces the Bekenstein--Hawking
entropy of each constituent.
\end{abstract}

\maketitle

\allowdisplaybreaks[1]
\raggedbottom

\section{Introduction}
Exact stationary black-hole binaries probe the conditions for equilibrium
in the fully nonlinear regime of general relativity. A regular equilibrium
in an axisymmetric binary requires smooth horizons and a smooth rotation
axis. A conical excess between the horizons signals a supporting strut.
Under asymptotically flat boundary conditions, nonexistence theorems rule
out stationary equilibrium for two black holes on a common axis in
vacuum gravity, including configurations with degenerate horizons
\cite{VacuumBalance,VacuumBalanceDegenerate}.

In Einstein--Maxwell theory, electrostatic repulsion can balance
gravitational attraction. The Majumdar--Papapetrou solutions realize this
equilibrium for extremal black holes with like-signed charges
\cite{MP,*Papapetrou1947}. Bonnor obtained an exact magnetostatic dipole
solution \cite{Bonnor1966}. Emparan interpreted it as an extremal dihole
with opposite magnetic charges. When both outer axis segments are
regular, a strut separates the horizons \cite{Emparan}.
The nonextremal Emparan--Teo dihole also has a strut \cite{EmparanTeo}.
A constant electromagnetic duality rotation gives a static dihole both
electric and magnetic charges while preserving its conicities
\cite{StaticDyonicDiholes}.
The stationary dihole of Manko, Rabad\'an and Sanabria-G\'omez retains a
strut between counterrotating Kerr--Newman constituents with opposite
electric charges \cite{StationaryDiholes}. A separate obstruction to
axis regularity is the Dirac--Misner string that connects constituents
with opposite NUT charges in Cl\'ement's rotating magnetized diholes
\cite{RotatingMagnetizedDiholes}.

An external background can change the axis conditions. Emparan used a
Harrison transformation to embed the Bonnor dihole in a Melvin magnetic
field and tuned the field strength to remove the strut \cite{Emparan}.
An axial Ehlers transformation generates a gravitational swirling
background with frame dragging that reverses across the equatorial plane
\cite{swirl}. Combining Ehlers and Harrison transformations also yields
a single dyonic Kerr--Newman black hole in a Melvin--swirling background
\cite{KNswirl}. For the binary problem, Astorino and Torresan showed that
the axial Ehlers map preserves the conicities of a static
double-Schwarzschild seed. They obtained a balanced vacuum binary from a
counterrotating extremal double-Kerr seed \cite{doubleKerr}. Here we ask
whether the electromagnetic field of a static magnetic dihole allows the
same map to remove its strut.

We apply the axial Ehlers transformation to the magnetostatic Bonnor
seed in the Ernst formulation of the Einstein--Maxwell equations. We
choose the seed gauge in which the azimuthal potential $A_\phi=A_m$
vanishes on both outer axis segments. Along the axis, the seed Ernst
potential reduces to $\mathcal E_0=-A_m^2$ and is nonzero only on the
inner segment. This electromagnetic term lets the map change the inner
conicity while preserving both outer conicities. Tuning the swirl
strength makes all three segments regular with a common $2\pi$ angular
period and a noncompact time coordinate. The balanced family has two
continuous parameters. Its two extremal horizons have compact
cross-sections and carry opposite angular momenta and opposite electric
and magnetic charges. Table~\ref{tab:equilibrium-comparison} compares
this construction with related binary solutions.
\begin{table*}[t]
\caption{Selected binary solutions and their axis structure.}
\label{tab:equilibrium-comparison}
\centering
\small
\setlength{\tabcolsep}{4pt}
\renewcommand{\arraystretch}{1.15}
\begin{tabular}{@{}p{.20\linewidth}p{.25\linewidth}p{.18\linewidth}p{.30\linewidth}@{}}
\toprule
Construction & Seed or method & Background & Axis and horizon properties \\
\midrule
Static dyonic dihole \cite{StaticDyonicDiholes}
& Duality rotation of the electrostatic Emparan--Teo dihole
& Same as the seed
& \raggedright The rotation changes the charges and preserves the seed conicities. \tabularnewline
Stationary dihole \cite{StationaryDiholes}
& Ernst--Manko--Ruiz solution
& Asymptotically flat
& \raggedright The counterrotating, oppositely charged pair retains a strut. \tabularnewline
Magnetized dihole \cite{Emparan}
& Harrison map of the magnetic Bonnor seed
& Melvin magnetic field
& \raggedright A tuned magnetic field removes the strut between the static constituents. \tabularnewline
Vacuum swirling binary \cite{doubleKerr}
& Axial Ehlers map of counterrotating double Kerr
& Vacuum swirling geometry
& \raggedright A tuned swirl balances two counterrotating extremal Kerr horizons. \tabularnewline
Present family
& Axial Ehlers map of the magnetic Bonnor seed
& Gravitational swirl
& \raggedright A tuned swirl balances the axis and the horizons carry opposite spins and dyonic charges. \tabularnewline
\bottomrule
\end{tabular}
\end{table*}

We give the metric and electromagnetic potentials in closed form. A
global time function excludes closed causal curves throughout the
exterior. Regular coordinates and gauge patches extend the metric and
Maxwell field smoothly across both degenerate Killing horizons. We
compute the horizon charges and verify the Smarr relation for the
Komar--Maxwell horizon mass with the chosen stationary generator and
electromagnetic potential references. The first law holds for variations
along the balanced family. The near-horizon metric and Maxwell field of
each constituent match those of extremal dyonic Kerr--Newman. The matched
mass parameter equals the isolated-horizon mass given by the dyonic
Christodoulou--Ruffini relation. Applying Kerr/CFT to each throat
reproduces its Bekenstein--Hawking entropy.

Sections~\ref{sec:generation} and \ref{sec:balance} derive the solution
and establish axis balance and exterior regularity.
Sections~\ref{sec:mechanics} and \ref{sec:near-horizon} present the horizon
mechanics, the near-horizon matching and the entropy calculation.
Appendices~\ref{app:omega} and \ref{app:regular-charts} give the field
reconstruction and the regular horizon charts and gauge patches.

\section{Bonnor black holes in the swirling universe}\label{sec:binary}
\subsection{Generation of the extremal solution}\label{sec:generation}
We use signature $(-,+,+,+)$ and units $G=c_{\rm light}=\hbar=1$. For a real
electromagnetic field $F=\dd A$ with $\star\star F=-F$, the Einstein field
equation is
\begin{equation}\label{eq:Einstein}
R_{ab}=2\left(F_{ac}F_b{}^c-\tfrac14g_{ab}F_{cd}F^{cd}\right).
\end{equation}
The Maxwell equations are
\begin{equation}\label{eq:Maxwell}
\dd F=0,\qquad \dd\star F=0.
\end{equation}
For stationary axisymmetric fields inheriting both symmetries, we use the
axial Lewis--Weyl--Papapetrou form
\begin{equation}\label{eq:axialmetric}
g=-f(\dd\phi-\omega\dd t)^2+
f^{-1}\left[\rho^2\dd t^2-e^{2\gamma_a}
(\dd\rho^2+\dd z^2)\right],
\end{equation}
where $f=-g_{\phi\phi}<0$. The azimuthal coordinate $\phi$ has period
$2\pi$ and time is noncompact. The exterior
orientation is $\dd t\wedge\dd\rho\wedge\dd z\wedge\dd\phi$.
Writing $m=\partial_\phi$, we introduce the scalars $\chi$ and $\psi$ by
$i_mF=-\dd\chi$ and $i_m\star F=-\dd\psi$. The electromagnetic and
gravitational Ernst potentials are
\begin{equation}\label{eq:Ernstdef}
\Phi=\chi+i\psi,\qquad \mathcal E=f-|\Phi|^2+ih.
\end{equation}
With twist potential $h$, the axial twist one-form is
\begin{equation}\label{eq:axial-twist}
\Theta=\dd h+2(\chi\dd\psi-\psi\dd\chi).
\end{equation}
An asterisk denotes complex conjugation, so $\Phi^*=\chi-i\psi$.
The reduced equations are
\begin{align}
f\Delta\mathcal E&=(\nabla\mathcal E+2\Phi^*\nabla\Phi)
\cdot\nabla\mathcal E,\notag\\
f\Delta\Phi&=(\nabla\mathcal E+2\Phi^*\nabla\Phi)
\cdot\nabla\Phi,
\label{eq:Ernst}
\end{align}
where $f=\operatorname{Re}\mathcal E+|\Phi|^2$.
Here $\Delta=\partial_\rho^2+\rho^{-1}\partial_\rho+\partial_z^2$
and $\nabla$ refer to the auxiliary flat cylindrical operators.
In these conventions the twist reconstruction is
\begin{equation}\label{eq:twist-rhoz}
\Theta_\rho=-\frac{f^2}{\rho}\omega_z,\qquad
\Theta_z=\frac{f^2}{\rho}\omega_\rho.
\end{equation}
The axial Ehlers symmetry acts as \cite{KNswirl}
\begin{equation}\label{eq:Ehlers}
\mathcal E' =\frac{\mathcal E}{1+ij\mathcal E},\;
\Phi'=\frac{\Phi}{1+ij\mathcal E},\;
f'=\frac{f}{|1+ij\mathcal E|^2}.
\end{equation}
The map leaves the reduced base function $\gamma_a$ unchanged.
Minkowski spacetime provides a simple example of the transformation.
For the flat seed with $f_0=\mathcal E_0=-\rho^2$, $\Phi_0=0$ and
$e^{2\gamma_a}=\rho^2$, the map gives \cite{swirl}
\begin{equation}
f'=-\frac{\rho^2}{1+j^2\rho^4},\qquad \omega'=4jz,
\end{equation}
after choosing the equatorial frame. The resulting vacuum background is
\begin{equation}\label{eq:swirl}
\begin{aligned}
g_{\rm swirl}&=(1+j^2\rho^4)(-\dd t^2+\dd\rho^2+\dd z^2)
\\
&\quad+\frac{\rho^2}{1+j^2\rho^4}(\dd\phi-4jz\dd t)^2.
\end{aligned}
\end{equation}

For the binary seed, introduce prolate spheroidal coordinates $(x,y)$ by
\begin{equation}\label{eq:prolate}
\rho=\sigma\sqrt{(x^2-1)(1-y^2)},\; z=\sigma xy,
\; \sigma>0.
\end{equation}
We take $x\geq1$ and $-1\leq y\leq1$. The constant $\sigma$ sets the
focal scale. The two foci lie at $(\rho,z)=(0,\pm\sigma)$, corresponding
to $(x,y)=(1,\pm1)$.
In Weyl coordinates, each horizon corresponds to one of these foci.
The orbit metric obeys
\begin{equation}
\dd\rho^2+\dd z^2=\sigma^2(x^2-y^2)
\left(\frac{\dd x^2}{x^2-1}+\frac{\dd y^2}{1-y^2}\right).
\end{equation}
To specify Bonnor's magnetic seed \cite{Bonnor1966}, interpreted as an
extremal dihole in Ref.~\cite{Emparan}, introduce a
dimensionless seed parameter $0<c<1$ and write $b=\sqrt{1-c^2}$.
The functions defining the seed solution are
\begin{equation}\label{eq:seedfunctions}
\begin{gathered}
D=x^2-1,\quad s=1-y^2,\quad p=D+b^2s,\\
\begin{alignedat}{2}
d&=(x+c)^2-b^2y^2, &\quad n&=p/d,\\
G&=\left(\frac p{x^2-y^2}\right)^4, &\quad
A_m&=-\frac{2\sigma cb(x+c)s}{p},\\
X&=\frac{\sigma^2Ds}{n^2}, &\quad S&=X+A_m^2.
\end{alignedat}
\end{gathered}
\end{equation}
The seed electromagnetic potential is $A_0=A_m\dd\phi$, and the seed metric is
\begin{equation}\label{eq:seedmetric}
g_0=-n^2\dd t^2+n^{-2}\left[G(\dd\rho^2+\dd z^2)
+\rho^2\dd\phi^2\right].
\end{equation}
The corresponding axial data are
\begin{equation}\label{eq:seedErnst}
f_0=-X,\; \Phi_0=A_m,\; \mathcal E_0=-S,\;
e^{2\gamma_a}=\frac{\rho^2G}{n^4}.
\end{equation}
The focal scale $\sigma>0$ and the dimensionless parameter $0<c<1$ specify
the extremal magnetic binary.
The individual magnetic fluxes are opposite, while the total magnetic
charge vanishes.
Expanding the seed fields at large $r=\sigma x$, with $y=\cos\theta$, gives
\begin{equation}\label{eq:seedasymptotic}
\begin{aligned}
(g_0)_{tt}&=-1+\frac{4c\sigma}{r}+O(r^{-2}),\\
A_m&=-\frac{2\sigma^2cb\sin^2\theta}{r}+O(r^{-2}).
\end{aligned}
\end{equation}
Thus the asymptotically flat seed mass is $2c\sigma$, its angular
momentum is zero, and its magnetic dipole moment in this orientation is
$-2\sigma^2cb$.

Applying the Ehlers map \eqref{eq:Ehlers} to the seed data
\eqref{eq:seedErnst} yields
\begin{equation}\label{eq:transformedErnst}
\mathcal E=-\frac S{1-ijS},\qquad \Phi=\frac{A_m}{1-ijS}.
\end{equation}
The real functions needed for reconstruction are
\begin{equation}\label{eq:newpot}
\begin{gathered}
H=1+j^2S^2,\qquad f=-\frac XH,\\
\chi=\frac{A_m}{H},\quad \psi=\frac{jSA_m}{H},\quad
h=-\frac{jS^2}{H}.
\end{gathered}
\end{equation}
Before applying the Ehlers transformation, we fix the additive constant
in $A_m$ by requiring $A_m=0$ on both outer axis segments.
Using $H=1+j^2S^2$ and $\psi=jS\chi$, we obtain
\[
\begin{aligned}
\dd h&=-\frac{2jS}{H^2}\dd S,\\
\chi\dd\psi-\psi\dd\chi
&=j\chi^2\dd S=\frac{jA_m^2}{H^2}\dd S.
\end{aligned}
\]
Substituting these expressions into the definition of $\Theta$ in
\eqref{eq:axial-twist} and using $S-A_m^2=X$ gives
\begin{equation}\label{eq:cancel}
\Theta=-\frac{2jX}{H^2}\dd S.
\end{equation}
With $f=-X/H$, the twist reconstruction equations \eqref{eq:twist-rhoz}
then give the Weyl-coordinate derivatives
\[
\omega_\rho=-\frac{2j\rho}{X}S_z,\qquad
\omega_z=\frac{2j\rho}{X}S_\rho.
\]
Applying the chain rule with the prolate spheroidal coordinates in
\eqref{eq:prolate} yields
\begin{equation}\label{eq:omegaPDE}
\omega_x=-\frac{2j\sigma s}{X}S_y,\qquad
\omega_y=\frac{2j\sigma D}{X}S_x.
\end{equation}
The generated metric is
\begin{equation}\label{eq:metric}
\begin{aligned}
g={}&-Hn^2\dd t^2+\frac XH(\dd\phi-\omega\dd t)^2\\
&\quad+\frac{HG\sigma^2(x^2-y^2)}{n^2}
\left(\frac{\dd x^2}{D}+\frac{\dd y^2}{s}\right).
\end{aligned}
\end{equation}
The electromagnetic potentials are
\begin{equation}\label{eq:fullA}
\begin{aligned}
A&=(U-\omega\chi)\dd t+\chi\dd\phi,\\
\widetilde A&=(\widetilde U-\omega\psi)\dd t+\psi\dd\phi.
\end{aligned}
\end{equation}
Appendix~\ref{app:omega} derives the rotation function $\omega$ and the
electromagnetic functions $U$ and $\widetilde U$ that enter the time
components of $A$ and $\widetilde A$.

Before imposing equilibrium, the generated family is parametrized by
$(\sigma,c,j)$. The axis conditions derived in the next subsection
determine $j$ in terms of the two seed parameters.
At large distance the rotation primitive has the leading behaviour
$\omega=4jz+O(1)$.
\subsection{Axis balance and exterior regularity}
\label{sec:balance}
The rotation axis consists of three segments, one inner and two outer.
On $\rho=0$, the inner segment has $-\sigma<z<\sigma$ ($x=1$, $|y|<1$).
The upper outer segment has $z>\sigma$ ($y=1$, $x>1$), and the lower outer
segment has $z<-\sigma$ ($y=-1$, $x>1$).
For a trial period $2\pi\Delta_\phi$, the normalized conicity is
\begin{equation}\label{eq:conicity}
\mathcal C=\frac{1}{2\pi}\lim_{\rho\to0}\frac{L}{R}
=\Delta_\phi\lim_{\rho\to0}
\frac{\sqrt{g_{\phi\phi}}}{\rho\sqrt{g_{\rho\rho}}}
=\left.\frac{\Delta_\phi}{H\sqrt G}\right|_{\rm axis}.
\end{equation}
Here $L$ is the circumference of a small circle around the axis and $R$
its proper radius. A segment is free of conical singularities when
$\mathcal C=1$.
On the outer axes $A_m=X=0$ and $G=H=1$. On the inner axis, denoted by
the subscript $A$,
\begin{equation}\label{eq:axisvalues}
\begin{aligned}
X&=0, & A_{m,A}&=-\frac{2\sigma c(1+c)}b,\\
G_A&=b^8, & H_A&=1+j^2A_{m,A}^4.
\end{aligned}
\end{equation}
For the seed ($j=0$), $\mathcal C_{\rm out}^{(0)}=\Delta_\phi$ and
$\mathcal C_A^{(0)}=\Delta_\phi b^{-4}$. Since $0<b<1$, setting each
conicity to unity requires a different $\Delta_\phi$, leaving an
unavoidable conical singularity in the seed.
For the transformed solution, the outer axes have $H\sqrt G=1$ and hence
fix $\Delta_\phi=1$. Regularity on the inner axis then requires
$H_A\sqrt{G_A}=1$. With $G_A=b^8$, this sets $H_A=b^{-4}$.
Substituting the inner-axis values from \eqref{eq:axisvalues} gives
\begin{equation}\label{eq:balance}
j=\eps\,\frac{\sqrt{2-c^2}}{4c(1+c)^2\sigma^2},\qquad
\Delta_\phi=1,\qquad \eps=\pm1.
\end{equation}
The balance condition shows that the electromagnetic term in the axial
Ernst potential is essential for the Ehlers transformation to remove the
conical singularities of the Bonnor seed.
The balance condition requires $j\neq0$ for $\sigma>0$ and $0<c<1$.
The two signs reverse the generated electric charge and spin. For each
fixed sign $\eps$, the balanced family has two continuous
parameters, $\sigma$ and $c$. All three axis segments share the same
$2\pi$-periodic angle. From here on,
physical horizon data refer to this balanced family.

In the off-axis exterior $x>1$, $|y|<1$, we have $D>0$ and $s>0$.
Thus $p=D+b^2s>0$, $d=p+2cx+2c^2>0$, and $H=1+j^2S^2\geq1$.
The $(t,\phi)$ block of the metric \eqref{eq:metric} has components
\[
g_{tt}=-Hn^2+\frac{X}{H}\omega^2,\qquad
g_{t\phi}=-\frac{X}{H}\omega,\qquad
g_{\phi\phi}=\frac{X}{H}.
\]
Its determinant is
\begin{equation}
g_{tt}g_{\phi\phi}-g_{t\phi}^{\,2}
=-n^2X=-\sigma^2Ds=-\rho^2<0.
\end{equation}
This block therefore has one negative and one positive eigenvalue.
The diagonal $(x,y)$ block has two positive coefficients, so the
four-dimensional metric has signature $(-+++)$.
In the same region, the local electromagnetic potential is
\[
A=(U-\omega\chi)\dd t+\chi\dd\phi,
\qquad \chi=\frac{A_m}{H}.
\]
The expressions for $U$ and $\omega$ are given in
Appendix~\ref{app:omega}, and their only coordinate-dependent denominator
is $d$. The denominator of $\chi$ is strictly positive in the off-axis
exterior. Thus the local coefficients of the metric and $A$ are real
analytic at every finite off-axis point, and so is the Maxwell field
$F=\dd A$.

Away from the foci, the metric near each axis segment takes the
Weyl-coordinate form
\[
g=-Hn^2\dd t^2+\frac{\rho^2}{Hn^2}(\dd\phi-\omega\dd t)^2
+\frac{HG}{n^2}(\dd\rho^2+\dd z^2).
\]
The functions $H,n,G,\omega$ are analytic in $\rho^2$ and $z$, with
$Hn^2>0$. For $\rho>0$,
\[
\frac{g_{\phi\phi}}{\rho^2}-g_{\rho\rho}=\frac{1-H^2G}{Hn^2},
\qquad g_{t\phi}=-\frac{\rho^2\omega}{Hn^2}.
\]
Axis balance gives $H(0,z)^2G(0,z)=1$ on each segment, so
$1-H^2G=O(\rho^2)$. Since $\phi$ is undefined on the axis, we examine
regularity in transverse Cartesian coordinates
$(\xi,\eta)=(\rho\cos\phi,\rho\sin\phi)$, for which
\[
\dd\rho^2+\rho^2\dd\phi^2=\dd\xi^2+\dd\eta^2,
\qquad \rho^2\dd\phi=\xi\dd\eta-\eta\dd\xi.
\]
The full metric then reads
\[
\begin{aligned}
g={}&\left(-Hn^2+\frac{\rho^2\omega^2}{Hn^2}\right)\dd t^2
\\
&\quad+\frac{HG}{n^2}(\dd\xi^2+\dd\eta^2+\dd z^2)
-\frac{2\omega}{Hn^2}\dd t\,(\xi\dd\eta-\eta\dd\xi)\\
&\quad+K(\rho^2,z)(\xi\dd\eta-\eta\dd\xi)^2,
\end{aligned}
\]
where $K(\rho^2,z)=(1-H^2G)/(\rho^2 Hn^2)$ for $\rho>0$ and extends
analytically to each axis segment away from the foci. The Cartesian chart
is regular near these axis segments. Appendix~\ref{app:regular-charts}
gives charts that extend across the extremal horizons, as well as the
electromagnetic gauge patches.
Throughout the exterior, including the regular axes,
\begin{equation}\label{eq:exterior-time}
g^{tt}=-\frac{1}{Hn^2}<0.
\end{equation}
The single-valued coordinate $t$ is therefore a time function, which
excludes closed causal curves in the exterior.
\subsection{Charges, Smarr relation and the first law of thermodynamics}\label{sec:mechanics}
\begin{figure}[!hb]
\centering
\includegraphics{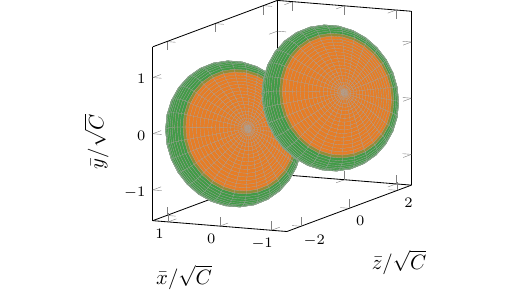}
\caption{Local isometric surface-of-revolution representations of the two
identical horizon metrics for $c=0.9$, in units $C=1$. Green and orange
denote Euclidean and pseudo-Euclidean meridional pieces, respectively.
The two surfaces are shifted apart for clarity; their plotted separation
does not represent a physical distance between the horizons.}
\label{fig:horizon-shape}
\end{figure}

To evaluate surface integrals on the upper horizon, introduce local
coordinates $\lambda>0$ and $-1\le u\le1$ near the upper focus by
\begin{equation}\label{eq:focus-main}
\rho=\lambda\sqrt{1-u^2},\qquad z=\sigma+\lambda u.
\end{equation}
Their exact relation to the prolate coordinates $(x,y)$ is
\begin{equation}\label{eq:focus-prolate}
x=\frac{R_-+\lambda}{2\sigma},\qquad
y=\frac{R_--\lambda}{2\sigma},
\end{equation}
where $R_-=\sqrt{\rho^2+(z+\sigma)^2}
=\sqrt{4\sigma^2+4\sigma\lambda u+\lambda^2}$.
In these local coordinates,
\[
\dd\rho^2+\dd z^2=\dd\lambda^2+
\frac{\lambda^2}{1-u^2}\dd u^2.
\]
For $\lambda>0$, the two-surfaces of constant $t$ and $\lambda$ have
$\dd t=\dd\lambda=0$. Their induced metric is
\[
q_\lambda=\frac{HG}{n^2}\frac{\lambda^2}{1-u^2}\dd u^2
+\frac{X}{H}\dd\phi^2.
\]
Taking the exterior limit $\lambda\to0^+$ gives the intrinsic metric of
a horizon cross-section,
\begin{equation}\label{eq:horizonmetric}
q=\lim_{\lambda\to0^+}q_\lambda
=\frac{\Gamma(u)}{1-u^2}\dd u^2+\gamma(u)\dd\phi^2,
\end{equation}
where the angular metric functions on the balanced branch are
\begin{equation}\label{eq:horizonfunctions}
\begin{gathered}
\Gamma(u)=C(1-c^2/2+c^2u^2/2),\\
C=\sigma^2c^2(1+c)^2,\qquad
\gamma(u)=\frac{C^2(1-u^2)}{\Gamma(u)}.
\end{gathered}
\end{equation}
Both poles are smooth with period $2\pi$, since $\Gamma(\pm1)=C$.
The intrinsic metric \eqref{eq:horizonmetric} also shows how the horizon
shape varies with $c$. The Gaussian curvature at the poles is
\[
K_H(\pm1)=\frac{1-2c^2}{C}.
\]
For $c>1/\sqrt{2}$, the curvature is negative near both poles, precluding
a global isometric embedding as a surface of revolution in Euclidean
three-space. Figure~\ref{fig:horizon-shape} illustrates the horizon geometry
for $c=0.9$ using Euclidean and pseudo-Euclidean pieces.

For the charge and angular-momentum integrals, we pull back the relevant
two-forms to surfaces of constant $t$ and $\lambda$ and take the same
exterior limit $\lambda\to0^+$. Expanding \eqref{eq:focus-prolate} at fixed
$u$ gives
\[
x=1+\frac{1+u}{2\sigma}\lambda+O(\lambda^2),\qquad
y=1-\frac{1-u}{2\sigma}\lambda+O(\lambda^2).
\]
Writing $L_u=2-c^2+c^2u$, the seed functions therefore satisfy
\[
p=\frac{L_u}{\sigma}\lambda+O(\lambda^2),\qquad
d=2c(1+c)+O(\lambda).
\]
The upper-horizon limits of the seed functions are
\[
A_{m,H}=-\frac{2\sigma cb(1+c)(1-u)}{L_u},\qquad
X_H=\frac{4C(1-u^2)}{L_u^2}.
\]
Thus
\[
S_H=X_H+A_{m,H}^2=\frac{4C(1-u)}{L_u}.
\]
The transformed horizon scalars are
\begin{equation}\label{eq:horizonpotentials}
\begin{aligned}
\chi_H&=\frac{A_{m,H}}{1+j^2S_H^2},\\
\psi_H&=\frac{jS_HA_{m,H}}{1+j^2S_H^2}.
\end{aligned}
\end{equation}
At the outer pole $u=1$, both $A_{m,H}$ and $S_H$ vanish.
At the inner pole $u=-1$, using the balance condition \eqref{eq:balance}
gives
\[
\begin{aligned}
\chi_H(-1)&=-2\sigma c(1+c)b^3,\\
\psi_H(-1)&=-8j\sigma^3c^3(1+c)^3b.
\end{aligned}
\]
From the definitions $i_mF=-\dd\chi$ and $i_m\star F=-\dd\psi$,
with $m=\partial_\phi$, we have
\[
F\big|_H=\chi_H'(u)\dd u\wedge\dd\phi,\qquad
(\star F)\big|_H=\psi_H'(u)\dd u\wedge\dd\phi.
\]
With the outward orientation $\dd u\wedge\dd\phi$ and angular period
$2\pi$, the upper-horizon charges are
\begin{align}
Q_m&=\frac1{4\pi}\int_H F
=\frac12\int_{-1}^{1}\chi_H'(u)\dd u
=\sigma c(1+c)b^3,\notag\\
Q_e&=\frac1{4\pi}\int_H\star F
=\frac12\int_{-1}^{1}\psi_H'(u)\dd u
=4c^3(1+c)^3bj\sigma^3.
\label{eq:charges}
\end{align}
The reflection $(y,\phi)\mapsto(-y,-\phi)$ preserves the metric and
reverses the Maxwell field. Consequently, the lower horizon carries
charges $-Q_m$ and $-Q_e$.

\begin{figure}[htbp]
\centering
\includegraphics{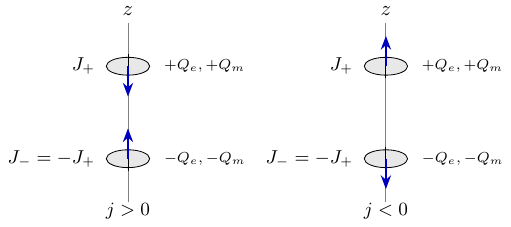}
\caption{The equilibrium branches with $j>0$ (left) and $j<0$ (right).
The arrows indicate angular momenta relative to the common axial
orientation. The charge labels give the upper-horizon values and
their negatives on the lower horizon.}
\label{fig:spin-branches}
\end{figure}

Smooth electromagnetic gauge representatives at ordinary axes and at the
horizons are discussed in
Appendix~\ref{app:EM-patches}.
The Komar--Maxwell angular momentum is
\[
J=\frac1{16\pi}\int_H\star\dd m^\flat
+\frac1{8\pi}\int_H
\left(\chi_H\,\star F-\psi_H\,F\right),\;
m=\partial_\phi.
\]
The gravitational term on the horizon is
\[
(\star\dd m^\flat)\big|_H
=\frac{kC\gamma}{\Gamma}\dd u\wedge\dd\phi,
\]
where
\begin{equation}\label{eq:rotation-parameter}
k=-4jC=-\eps c\sqrt{2-c^2}.
\end{equation}
Using the electromagnetic pullbacks, the horizon scalars and the balance
condition \eqref{eq:balance}, integration over the $2\pi$-periodic angle gives
\begin{equation}\label{eq:Jintegral}
\begin{aligned}
J&=\int_{-1}^{1}\left[
\frac{kC\gamma}{8\Gamma}
+\frac{\chi_H\psi_H'-\psi_H\chi_H'}4\right]\dd u
\\
&=\left[-\frac{jC^2[2u+b^2(1-u^2)]}{2-c^2+c^2u^2}\right]_{-1}^{1}
=-2jC^2.
\end{aligned}
\end{equation}
The same reflection reverses the axial generator and gives opposite
angular momenta for the two horizons:
\begin{equation}\label{eq:paircharges}
J_+=J,\qquad J_-=-J.
\end{equation}
Figure~\ref{fig:spin-branches} shows the two equilibrium branches.

The metric coefficients near $\lambda=0$ have the expansions
\[
\begin{gathered}
Hn^2=\frac{\Gamma}{C^2}\lambda^2+O(\lambda^3),\\
\frac{HG}{n^2}=\frac{\Gamma}{\lambda^2}+O(\lambda^{-1}),\qquad
\frac XH=\gamma+O(\lambda).
\end{gathered}
\]
For the upper horizon, write $\ell=T+\Omega m$ with $T=\partial_t$.
The metric \eqref{eq:metric} gives
\[
g(\ell,\ell)=-Hn^2+\frac{X}{H}(\Omega-\omega)^2.
\]
Since $Hn^2\to0$ and $X/H\to\gamma>0$ away from the poles,
the null condition and the expansion of \eqref{eq:omega} give
\begin{equation}\label{eq:Omega}
\Omega=\lim_{\lambda\to0^+}\omega=4j\sigma(1-2c).
\end{equation}
The same expansion gives $\omega-\Omega=4j\lambda+O(\lambda^2)$.
In the regular horizon chart of Appendix~\ref{app:regular-charts},
\[
g(\ell,\ell)=O(\lambda^2),\qquad
\left.\dd(\ell^2)\right|_H=-2\kappa\left.\ell^\flat\right|_H=0.
\]
The surface gravity and Hawking temperature therefore vanish,
\begin{equation}\label{eq:temperature}
\kappa=T_H=0.
\end{equation}
The reflection $(y,\phi)\mapsto(-y,-\phi)$ maps $\ell$ to $T-\Omega m$
on the lower horizon, where the surface gravity and temperature also vanish.
Since $\sqrt{\det q}=C$, the horizon areas are
\begin{equation}\label{eq:area}
\Area_\pm=\int_0^{2\pi}\dd\phi\int_{-1}^{1}C\,\dd u=4\pi C,
\end{equation}
and the corresponding entropies are $S_{H,\pm}=\Area_\pm/4=\pi C$.
The flux and spin integrals give
\begin{equation}\label{eq:extreme-area}
\Area_\pm=4\pi\sqrt{4J_\pm^2+Q^4},
\end{equation}
where $Q^2:=Q_e^2+Q_m^2=C(1-c^2)$.
Each horizon thus satisfies the area--angular momentum--charge equality
of extremal Kerr--Newman.

For the upper component, the expansions
\eqref{eq:U-horizon-expansion}--\eqref{eq:Ud-horizon-expansion} give the
constant horizon potentials
\begin{equation}\label{eq:timepotentials}
\begin{aligned}
U_H&=(A_t+\Omega\chi)\big|_H=-4cbj\sigma^2,\\
\widetilde U_H&=(\widetilde A_t+\Omega\psi)\big|_H=\frac b{1+c}.
\end{aligned}
\end{equation}
The Komar--Maxwell charge two-form associated with $T$ is
\begin{equation}\label{eq:mass-current}
\mathcal Q_T=\frac{-1}{8\pi}
\left(\star\dd T^\flat+2A_t\star F-2\widetilde A_tF\right),
\, \dd\mathcal Q_T=0.
\end{equation}
The Komar--Maxwell horizon mass associated with $T=\partial_t$ is
$M_H^{(t)}=\int_H\mathcal Q_T$.
We use the equatorial references $U(x,0)=\widetilde U(x,0)=0$ of
Appendix~\ref{app:omega}.
With $T=\ell-\Omega m$ and the potentials in \eqref{eq:fullA},
the gravitational and electromagnetic terms combine to give
\begin{align*}
M_H^{(t)}
&=-\frac1{8\pi}\int_H
\left(\star\dd\ell^\flat+2U_H\star F-2\widetilde U_HF\right)
+2\Omega J\\
&=\frac{\kappa\Area}{4\pi}+2\Omega J-U_HQ_e+\widetilde U_HQ_m,
\end{align*}
where $\int_H\star\dd\ell^\flat=-2\kappa\Area$.
For $\kappa=0$, this gives
\[
M_H^{(t)}=c\sigma\left(b^4+16j^2C^2\right).
\]
Using the balance condition \eqref{eq:balance}, this reduces to
$M_H^{(t)}=c\sigma$.
To express the charge terms in Smarr form, choose the local gauge
representatives
\[
\bar A=A+Q_m\dd\phi,\qquad
\overline{\widetilde A}=\widetilde A+Q_e\dd\phi.
\]
Their horizon angular components are $\bar\chi=\chi_H+Q_m$ and
$\bar\psi=\psi_H+Q_e$. At each fixed solution these shifts leave $F$
and $\star F$ unchanged.
The horizon potentials evaluated with these representatives are therefore
\begin{equation}\label{eq:thermopotentials}
\begin{aligned}
\Phi_e&=-\bar A(\ell)\big|_H=-U_H-\Omega Q_m,\\
\Phi_m&=\overline{\widetilde A}(\ell)\big|_H
=\widetilde U_H+\Omega Q_e.
\end{aligned}
\end{equation}
The mass formula takes the Smarr form
\begin{equation}\label{eq:Smarr}
M_H^{(t)}=\frac{\kappa\Area}{4\pi}
+2\Omega J+\Phi_eQ_e+\Phi_mQ_m=c\sigma.
\end{equation}
Differentiating $M_H^{(t)}=c\sigma$ yields
\begin{equation}\label{eq:firstlaw}
\dd M_H^{(t)}=T_H\,\dd S_H+\Omega\,\dd J
+\Phi_e\,\dd Q_e+\Phi_m\,\dd Q_m,
\end{equation}
where $T_H=0$. Here $\dd$ is the differential on the balanced family with coordinates
$(\sigma,c)$ at fixed $\eps$, with $j=j(\sigma,c)$ given by
\eqref{eq:balance}. The same relation holds for the lower horizon by
reflection.

A constant electromagnetic duality rotation preserves the horizon area
and Komar--Maxwell angular momentum and brings the charges to a purely
electric representative with charge squared $Q^2=Q_e^2+Q_m^2$.
In the Kerr--Newman isolated-horizon normalization \cite{IH}, the
Christodoulou--Ruffini relation then gives
\begin{align}
M_{\rm IH}^2
&=\frac{S_H}{4\pi}+\frac{Q^2}{2}
+\frac{\pi(Q^4+4J^2)}{4S_H},\label{eq:CRmass}\\
&=\frac{(C+Q^2)^2+4J^2}{4C}
=C(1-c^2/2).\notag
\end{align}
Here $S_H=\pi C$. The masses $M_{\rm IH}$ and
$M_H^{(t)}$ are distinct. The next subsection identifies $M_{\rm IH}$
with the mass parameter of the extremal Kerr--Newman throat.
\subsection{Extremal Kerr--Newman matching and Kerr/CFT entropy}
\label{sec:near-horizon}
To magnify the upper extremal throat, introduce
\begin{equation}\label{eq:NHscaling}
\lambda=\eta r,\; t=C\tau/\eta,\;
\phi=\varphi+\Omega t,\; \eta\longrightarrow0.
\end{equation}
After the local gauge shifts $A\mapsto A-U_H\dd t$ and
$\widetilde A\mapsto\widetilde A-\widetilde U_H\dd t$ remove the divergent
terms, the near-horizon limit gives
\begin{align}
g_{\rm NH}&=\Gamma\left(-r^2\dd\tau^2+\frac{\dd r^2}{r^2}
+\frac{\dd u^2}{1-u^2}\right)\notag\\
&\quad+\gamma(\dd\varphi+kr\dd\tau)^2,\label{eq:NHmetric}\\
A_{\rm NH}&=\chi_H(\dd\varphi+kr\dd\tau),\notag\\
\widetilde A_{\rm NH}&=\psi_H(\dd\varphi+kr\dd\tau)
+C u_{d,1}r\dd\tau.
\label{eq:NHfields}
\end{align}
where $\varphi=\phi-\Omega t$ has period $2\pi$ and $u_{d,1}$ is given
in \eqref{eq:dual-radial-coefficient}.
The two-dimensional radial-time factor is of $AdS_2$ type. The lower throat
follows by reflection and has the opposite charges and spin.
The near-horizon scaling isolates each degenerate horizon from its companion
and the swirling outer end, so we carry out the asymptotic-symmetry analysis
separately for each throat.

We now compare the near-horizon metric \eqref{eq:NHmetric} and
electromagnetic one-forms \eqref{eq:NHfields} with those of an extremal
dyonic Kerr--Newman throat. Let $M_0$, $a_0$, $Q_{e,0}$ and $Q_{m,0}$
denote the Kerr--Newman parameters, with the extremality relation
$M_0^2=a_0^2+Q_{e,0}^2+Q_{m,0}^2$. In terms of the polar coordinate
$u_0=\cos\theta$, the Kerr--Newman near-horizon metric
\cite{KNreview,NHreview,HartmanCFT} is
\begin{equation}\label{eq:KNmetric}
\begin{aligned}
g_{\rm KN}
&=\Gamma_0\left(-r^2\dd\tau^2+\frac{\dd r^2}{r^2}
+\frac{\dd u_0^2}{1-u_0^2}\right)\\
&\quad+\frac{C_0^2(1-u_0^2)}{\Gamma_0}
(\dd\varphi+k_0r\dd\tau)^2,
\end{aligned}
\end{equation}
where
\[
C_0=M_0^2+a_0^2,\quad
\Gamma_0=M_0^2+a_0^2u_0^2,\quad
k_0=\frac{2M_0a_0}{C_0}.
\]
For the orientation $-\dd\tau\wedge\dd r\wedge\dd u_0\wedge\dd\varphi$,
we take the zero-cosmological-constant limit of the potential in
Ref.~\cite{HartmanCFT}, with charge parameters $q_e=Q_{e,0}$ and
$q_m=-Q_{m,0}$. For the rotating throats considered here, the local
gauge shift $-(e_0/k_0)\dd\varphi$ gives the representative used below.
Replacing $(Q_{e,0},Q_{m,0})$ by $(-Q_{m,0},Q_{e,0})$ gives the dual
potential, with $\dd\widetilde A_{\rm KN}=\star\dd A_{\rm KN}$.
The two local potentials are
\begin{equation}\label{eq:KNpotentials}
\begin{aligned}
A_{\rm KN}&=f_0(u_0)(\dd\varphi+k_0r\dd\tau)+e_0r\dd\tau,\\
\widetilde A_{\rm KN}
&=\widetilde f_0(u_0)(\dd\varphi+k_0r\dd\tau)+\widetilde e_0r\dd\tau,
\end{aligned}
\end{equation}
with
\begin{equation}\label{eq:KNangular}
\begin{aligned}
f_0(u_0)&=\frac{-C_0Q_{m,0}u_0+Q_{e,0}M_0a_0(1-u_0^2)}{\Gamma_0},\\
\widetilde f_0(u_0)
&=\frac{-C_0Q_{e,0}u_0-Q_{m,0}M_0a_0(1-u_0^2)}{\Gamma_0},
\end{aligned}
\end{equation}
and
\begin{equation}\label{eq:KNconstants}
\begin{aligned}
e_0&=\frac{Q_{e,0}(M_0^2-a_0^2)}{C_0},\\
\widetilde e_0&=-\frac{Q_{m,0}(M_0^2-a_0^2)}{C_0}.
\end{aligned}
\end{equation}

The metric \eqref{eq:NHmetric} takes the Kerr--Newman form under the
identification
\begin{equation}\label{eq:KNmap}
\begin{gathered}
u_0=-u,\qquad M_0=\sqrt{C(1-c^2/2)},\\
a_0=\frac{J}{M_0}=-\eps\sqrt{Cc^2/2}.
\end{gathered}
\end{equation}
These relations give $C_0=C$, $\Gamma_0(-u)=\Gamma(u)$ and $k_0=k$.
Identifying $Q_{e,0}=Q_e$ and $Q_{m,0}=Q_m$ gives
\[
M_0^2-a_0^2=C(1-c^2)=Q_e^2+Q_m^2.
\]

To compare the angular components of \eqref{eq:NHfields} in the same
gauge, add $Q_m\dd\varphi$ to $A_{\rm NH}$ and $Q_e\dd\varphi$ to
$\widetilde A_{\rm NH}$. The resulting coefficients of $\dd\varphi$
are
\begin{equation}\label{eq:KNangular-match}
\begin{aligned}
\bar\chi&=\chi_H+Q_m=f_0(-u),\\
\bar\psi&=\psi_H+Q_e=\widetilde f_0(-u).
\end{aligned}
\end{equation}
The coefficients of $r\dd\tau$ in the shifted potentials are
$k\bar\chi-kQ_m$ and $k\bar\psi+C u_{d,1}-kQ_e$, respectively.
Axis balance gives
\begin{equation}\label{eq:KNradial}
\begin{aligned}
-kQ_m&=Q_e b^2=e_0,\\
C u_{d,1}-kQ_e&=-Q_m b^2=\widetilde e_0.
\end{aligned}
\end{equation}
These identities complete the matching of both one-forms.
The local uniqueness theorem for axisymmetric extremal isolated
horizons relates the induced horizon metric, rotation one-form and
Maxwell pullback to those of extremal Kerr--Newman
\cite{ExtremalIHUniqueness}. The theorem states its electromagnetic
result at zero magnetic charge. Each constituent admits a purely
electric representative by the duality rotation described in
Sec.~\ref{sec:mechanics}.

The corresponding Kerr--Newman horizon quantities are
\begin{equation}
\begin{gathered}
\Area_0=4\pi(M_0^2+a_0^2)=4\pi C,\\
J_0=M_0a_0=J,\qquad
M_0=M_{\rm IH}.
\end{gathered}
\end{equation}
The matched mass is $M_{\rm IH}$, not $M_H^{(t)}$.

Having matched the complete near-horizon fields and the angular period,
we adopt the standard Kerr/CFT boundary conditions for the extremal
Kerr--Newman throat \cite{HartmanCFT,CompereCentral}. Under these boundary
conditions, the Maxwell sector does not contribute to the Virasoro central charge
\cite{CompereCentral}.
For either constituent choose the local angular orientation along its
spin, and write $\kappa_{\rm rot}=|k|>0$. The period remains $2\pi$.
The central charge is
\begin{equation}\label{eq:CFTcentral}
c_L=3\kappa_{\rm rot}\int_{-1}^{1}
\sqrt{\frac{\Gamma\gamma}{1-u^2}}\,\dd u
=6\kappa_{\rm rot}C=12|J|.
\end{equation}
The rotational Frolov--Thorne temperature is
\cite[Sec.~5]{HartmanCFT}
\begin{equation}\label{eq:CFTtemperature}
T_L=\frac1{2\pi\kappa_{\rm rot}}=\frac{C}{4\pi|J|}.
\end{equation}
The microscopic degrees of freedom of the dual chiral CFT are counted
through the Cardy formula, which gives the entropy
\begin{equation}\label{eq:CFTentropy}
S_{\rm CFT}=\frac{\pi^2}{3}c_LT_L=\pi C=\frac{\Area}{4}.
\end{equation}
Within the Kerr/CFT correspondence, this gives the microscopic entropy
of each constituent and reproduces its Bekenstein--Hawking entropy.
The strict domain $0<c<1$ has $J\ne0$, satisfying the nonzero-spin
condition required by the rotational Kerr/CFT formulas used here.
\section{Conclusions}
An axial Ehlers transformation of the magnetostatic Bonnor seed yields
an exact two-parameter family of stationary Einstein--Maxwell equilibria
in a non-asymptotically flat swirling background. In the chosen seed
gauge, the magnetic contribution to the axial Ernst potential vanishes
on the outer axes and is nonzero on the inner segment. The swirl can
therefore tune the inner conicity while the outer axes stay regular.
The balance condition \eqref{eq:balance} removes the strut and makes
all three segments regular for a common $2\pi$ angular period with
noncompact time. This mechanism complements the Melvin-supported
magnetic dihole and the vacuum swirling double-Kerr binary in
Table~\ref{tab:equilibrium-comparison}.

The two extremal Killing horizons have compact cross-sections and
carry opposite angular momenta and opposite electric and magnetic
charges. Each horizon obeys the extremal area--angular momentum--charge
relation. Regular charts establish that the metric and Maxwell field
are smooth on the axis and extend smoothly across both horizons. A
global time function excludes closed causal curves throughout the
exterior, including the regular axes.

With the chosen stationary generator and electromagnetic potential
references, the positive Komar--Maxwell horizon mass $M_H^{(t)}$
satisfies the Smarr relation. Its variations along the balanced family
satisfy the first law. The near-horizon metric and Maxwell field of
each constituent match those of extremal dyonic Kerr--Newman, whose mass
parameter equals the isolated-horizon mass $M_{\rm IH}$ given by the
dyonic Christodoulou--Ruffini relation. The matching describes each
local throat within the full binary spacetime, which also contains
the companion and the swirling outer end. For each throat, the Kerr/CFT
central charge and Frolov--Thorne temperature reproduce the
Bekenstein--Hawking entropy through the Cardy formula.

The perturbative stability of the equilibria and the global extension
needed to identify event horizons remain open.

\acknowledgments
We thank Meng Xu for carefully reviewing the manuscript and helping to revise it.

This work is supported by the National Key Research and Development Program of China (No.~2021YFC2203001) and the National Natural Science Foundation of China (No.~12475049).

\appendix
\section{The rotation function \texorpdfstring{$\omega(x,y)$}{omega(x,y)} and the potential \texorpdfstring{$U(x,y)$}{U(x,y)}}\label{app:omega}
Differentiating $S=X+A_m^2$ in \eqref{eq:seedfunctions} gives
\[
\begin{aligned}
S_x&=\frac{2\sigma^2s}{p^2}
\bigl\{(x-2c)d^2\\
&\qquad+2cb^2\bigl[(1-3y^2)d+2b^2y^2s\bigr]\bigr\},\\
S_y&=-\frac{2\sigma^2Dy}{p^2}
\left[d^2-4cb^2(x+c)s\right].
\end{aligned}
\]
Substituting these derivatives into the first equation of
\eqref{eq:omegaPDE} gives
\[
\omega_x=4j\sigma y
\left[1-\frac{4cb^2(x+c)s}{d^2}\right].
\]
Since $\partial_x(d^{-1})=-2(x+c)/d^2$, integration with respect to $x$
yields
\[
\omega=4j\sigma xy+\frac{8j\sigma cb^2ys}{d}+\mathcal W(y).
\]
The second equation of \eqref{eq:omegaPDE} then requires
\[
\mathcal W'(y)=-8j\sigma c,\qquad
\mathcal W(y)=-8j\sigma cy+\omega_{\rm c}.
\]
The additive constant fixes the rotating frame. Since
$\omega(x,0)=\omega_{\rm c}$, the equatorial condition $\omega(x,0)=0$
sets $\omega_{\rm c}=0$. Expanding $d=(x+c)^2-b^2y^2$, we obtain
\begin{equation}\label{eq:omega}
\omega=\frac{4j\sigma y}{d}
\left(2c-4c^3-3c^2x+x^3-xy^2+c^2xy^2\right).
\end{equation}

The functions $U$ and $\widetilde U$ in \eqref{eq:fullA} follow from the
duality relations $F=\dd A$ and $\star F=\dd\widetilde A$.
Evaluating the Hodge dual with the metric \eqref{eq:metric} gives
\[
\begin{aligned}
(A_t)_x+\omega\chi_x&=\frac{\sigma sH}{X}\psi_y,
&
(A_t)_y+\omega\chi_y&=-\frac{\sigma DH}{X}\psi_x,\\
(\widetilde A_t)_x+\omega\psi_x&=-\frac{\sigma sH}{X}\chi_y,
&
(\widetilde A_t)_y+\omega\psi_y&=\frac{\sigma DH}{X}\chi_x.
\end{aligned}
\]
For $U=A_t+\omega\chi$ and $\widetilde U=\widetilde A_t+\omega\psi$,
differentiation introduces the additional terms $\chi\,\dd\omega$ and
$\psi\,\dd\omega$. With the potentials \eqref{eq:newpot}, the resulting
combinations simplify to
\[
\begin{aligned}
H\psi_i-2j\chi S_i&=j\left[S(A_m)_i-A_mS_i\right],\\
-H\chi_i-2j\psi S_i&=-(A_m)_i,
\end{aligned}
\]
where $i=x,y$. By \eqref{eq:omegaPDE}, the terms
$\chi\,\partial_i\omega$ and $\psi\,\partial_i\omega$ contribute
$-2j\chi S_i$ and $-2j\psi S_i$ to these combinations, respectively.
Hence
\[
\begin{gathered}
U_x=\frac{j\sigma s}{X}\left[S(A_m)_y-A_mS_y\right],\\
U_y=-\frac{j\sigma D}{X}\left[S(A_m)_x-A_mS_x\right],\\
\widetilde U_x=-\frac{\sigma s}{X}(A_m)_y,\quad
\widetilde U_y=\frac{\sigma D}{X}(A_m)_x.
\end{gathered}
\]

Substituting the seed functions, the $x$-equations reduce to
\[
U_x=-\frac{4cb^3j\sigma^2ys^2(x+c)}{d^2},\qquad
\widetilde U_x=-\frac{4cby(x+c)}{d^2}.
\]
Integration gives
\[
U=\frac{2cb^3j\sigma^2ys^2}{d}+\mathcal U(y),\qquad
\widetilde U=\frac{2cby}{d}+\mathcal V(y),
\]
while the remaining $y$-equations require
\[
\mathcal U'(y)=-6cbj\sigma^2(1-y^2),\qquad
\mathcal V'(y)=0.
\]
Choosing the electromagnetic reference constants by
$U(x,0)=\widetilde U(x,0)=0$ fixes
$\mathcal U(y)=-2cbj\sigma^2(3y-y^3)$ and $\mathcal V(y)=0$. Thus
\begin{equation}\label{eq:U}
\begin{aligned}
U&=-\frac{2cbj\sigma^2y}{d}
\bigl(-1+4c^2+6cx+3x^2\\
&\quad-y^2-2cxy^2-x^2y^2\bigr),
\end{aligned}
\end{equation}
and
\begin{equation}\label{eq:Ud}
\widetilde U=\frac{2cby}{d}.
\end{equation}
Together with \eqref{eq:newpot}, these functions determine the local
potentials $A$ and $\widetilde A$ in \eqref{eq:fullA}.

Near the upper horizon, expand \eqref{eq:focus-prolate} at fixed $u$:
\[
x=1+\frac{1+u}{2\sigma}\lambda+O(\lambda^2),\qquad
y=1+\frac{u-1}{2\sigma}\lambda+O(\lambda^2).
\]
Consequently,
\[
d=2c(1+c)+\frac{(1+c)(2-c+cu)}{\sigma}\lambda+O(\lambda^2).
\]
Direct substitution into \eqref{eq:U} gives
\begin{equation}\label{eq:U-horizon-expansion}
\begin{aligned}
U&=-2cbj\sigma^2
\frac{4c(1+c)+\dfrac{2(1+c)(2-c+cu)}{\sigma}\lambda+O(\lambda^2)}
{2c(1+c)+\dfrac{(1+c)(2-c+cu)}{\sigma}\lambda+O(\lambda^2)}\\
&=-4cbj\sigma^2+O(\lambda^2)=U_H+O(\lambda^2).
\end{aligned}
\end{equation}
Similarly, \eqref{eq:Ud} yields
\begin{equation}\label{eq:Ud-horizon-expansion}
\begin{aligned}
\widetilde U&=
\frac{2cb\left[1+\dfrac{u-1}{2\sigma}\lambda+O(\lambda^2)\right]}
{2c(1+c)+\dfrac{(1+c)(2-c+cu)}{\sigma}\lambda+O(\lambda^2)}\\
&=\frac b{1+c}-\frac b{\sigma c(1+c)}\lambda+O(\lambda^2)\\
&=\widetilde U_H+u_{d,1}\lambda+O(\lambda^2),
\end{aligned}
\end{equation}
where
\begin{equation}\label{eq:dual-radial-coefficient}
u_{d,1}=-\frac b{\sigma c(1+c)}.
\end{equation}
\section{Additional axis and horizon charts}\label{app:regular-charts}
\subsection{Discovering the regular coordinates}
Along the radial null directions $\dd u=0$ and
\mbox{$\dd\phi-\omega\dd t=0$}, the metric \eqref{eq:metric}, expressed in
the local coordinates \mbox{\eqref{eq:focus-main}--\eqref{eq:focus-prolate}}, gives
\[
-Hn^2\dd t^2+\frac{HG}{n^2}\dd\lambda^2=0.
\]
Hence $\dd t/\dd\lambda=\pm T_{\rm ex}$ with
\begin{equation}\label{eq:radialseries}
T_{\rm ex}:=\frac{\sqrt G}{n^2}
=\left(\frac d{x^2-y^2}\right)^2
=\frac C{\lambda^2}+\frac\alpha\lambda+O(1),
\end{equation}
where $\alpha=\sigma c(1+c)^2(2-c)$.
Here $C$ and $\Omega$ are defined in \eqref{eq:horizonfunctions} and
\eqref{eq:Omega}.
Let $T=C/\lambda^2+\alpha/\lambda$ and
$r_*=-C/\lambda+\alpha\log|\lambda|$, so $\dd r_*=T\dd\lambda$.
On the ingoing branch, $\dd t/\dd\lambda=-T_{\rm ex}$, so
$\dd(t+r_*)/\dd\lambda=T-T_{\rm ex}=O(1)$.
Along the same ingoing null curve, $\dd\phi=\omega\dd t$ and the
near-horizon expansion $\omega-\Omega=4j\lambda+O(\lambda^2)$ gives
\[
\frac{\dd(\phi-\Omega t)}{\dd\lambda}
=-(\omega-\Omega)T_{\rm ex}=-\frac{4jC}{\lambda}+O(1).
\]
Integrating gives $\phi-\Omega t=-4jC\log|\lambda|+O(1)$, so we define
\begin{equation}\label{eq:ingoing}
v=t+r_*,\qquad
\varphi_{\rm in}=\phi-\Omega t+4jC\log|\lambda|.
\end{equation}
Indeed
\begin{equation}
\dd\phi-\omega\dd t=
\dd\varphi_{\rm in}-(\omega-\Omega)\dd v+B_\lambda\dd\lambda,
\end{equation}
where
\[
B_\lambda=(\omega-\Omega)T-\frac{4jC}{\lambda}.
\]
The simple poles cancel, leaving $B_\lambda$ analytic.
Substituting \eqref{eq:ingoing} and the
preceding one-form identity into the metric gives
\[
\begin{aligned}
\dd s^2={}&-Hn^2(\dd v-T\dd\lambda)^2\\
&+\frac XH
\left[\dd\varphi_{\rm in}
-(\omega-\Omega)\dd v+B_\lambda\dd\lambda\right]^2\\
&+\frac{HG}{n^2}
\left(\dd\lambda^2+\frac{\lambda^2}{1-u^2}\dd u^2\right).
\end{aligned}
\]
The radial component is therefore
\begin{equation}\label{eq:radial-cancellation}
g_{\lambda\lambda}
=\frac{HG}{n^2}-Hn^2T^2+\frac XH B_\lambda^2.
\end{equation}
Using $T_{\rm ex}=\sqrt G/n^2$, the first two terms combine to
$Hn^2(T_{\rm ex}-T)(T_{\rm ex}+T)$.
The remainder $T_{\rm ex}-T$ in \eqref{eq:radialseries} is analytic at
$\lambda=0$, as are the combinations
\[
\lambda^2(T_{\rm ex}+T)=2C+O(\lambda),\qquad
\frac{Hn^2}{\lambda^2}=\frac{\Gamma}{C^2}+O(\lambda).
\]
Thus the difference of the first two terms is finite and analytic.
Since $X/H$ and $B_\lambda$ are analytic as well, the full component
$g_{\lambda\lambda}$ in \eqref{eq:radial-cancellation} is regular.
The mixed time--radial component is
\[
g_{v\lambda}
=Hn^2T-\frac XH(\omega-\Omega)B_\lambda.
\]
As $\lambda\to0$, the first term tends to $\Gamma/C$, whereas the
second vanishes because $\omega-\Omega=O(\lambda)$. Hence
\[
g_{v\lambda}\big|_{\lambda=0}=\frac{\Gamma(u)}{C}>0.
\]
Away from the polar coordinate singularities, the horizon determinant is
\begin{equation}\label{eq:horizon-nondegeneracy}
\det g\big|_{\lambda=0}=-\Gamma^2<0,
\end{equation}
so the ingoing chart defines a nondegenerate Lorentz metric across
$\lambda=0$. The horizon generator is $\ell=\partial_v=T+\Omega m$.
Since $g(\ell,\ell)=O(\lambda^2)$, its surface gravity vanishes, in
agreement with \eqref{eq:temperature}.

The outgoing chart is
\begin{equation}\label{eq:outgoing}
w=t-r_*,\qquad
\varphi_{\rm out}=\phi-\Omega t-4jC\log|\lambda|.
\end{equation}
In this chart,
\[
\dd\phi-\omega\dd t
=\dd\varphi_{\rm out}-(\omega-\Omega)\dd w-B_\lambda\dd\lambda.
\]
The component $g_{\lambda\lambda}$ is unchanged, while the mixed
component has the limit $g_{w\lambda}|_{\lambda=0}=-\Gamma/C$.

\subsection{Regular electromagnetic potentials and gauge patches}\label{app:EM-patches}
In the regular charts, write $t_{\rm in}=v$ and $t_{\rm out}=w$.
The potentials \eqref{eq:fullA} become
\[
\begin{aligned}
A_{\rm in/out}
&=[U-(\omega-\Omega)\chi]\,\dd t_{\rm in/out}\\
&\quad+\chi\,\dd\varphi_{\rm in/out}
\pm(-UT+\chi B_\lambda)\,\dd\lambda,\\
\widetilde A_{\rm in/out}
&=[\widetilde U-(\omega-\Omega)\psi]\,\dd t_{\rm in/out}\\
&\quad+\psi\,\dd\varphi_{\rm in/out}
\pm(-\widetilde U T+\psi B_\lambda)\,\dd\lambda,
\end{aligned}
\]
where the upper sign is ingoing and the lower sign is outgoing.
To obtain smooth gauges across the horizon, we first cancel the radial
poles by choosing the local representatives
\begin{equation}\label{eq:regular-EM-patches}
\widehat A_{\rm in/out}=A\mathbin{\pm}\dd\Lambda,\qquad
\widehat{\widetilde A}_{\rm in/out}
=\widetilde A\mathbin{\pm}\dd\widetilde\Lambda,
\end{equation}
with
\[
\dd\Lambda=U_HT\,\dd\lambda,\qquad
\dd\widetilde\Lambda=\left(\widetilde U_HT+\frac{Cu_{d,1}}{\lambda}\right)\dd\lambda.
\]
The shifted radial coefficients are
\begin{equation}\label{eq:regular-EM-radial}
\begin{aligned}
(\widehat A_{\rm in/out})_\lambda
&=\pm\left[-(U-U_H)T+\chi B_\lambda\right],\\
(\widehat{\widetilde A}_{\rm in/out})_\lambda
&=\pm\left[-(\widetilde U-\widetilde U_H)T
+\psi B_\lambda+\frac{Cu_{d,1}}\lambda\right].
\end{aligned}
\end{equation}
The expansion \eqref{eq:U-horizon-expansion} makes $(U-U_H)T$ analytic
at $\lambda=0$. For the dual potential,
\eqref{eq:Ud-horizon-expansion} gives
\[
(\widetilde U-\widetilde U_H)T
=\frac{Cu_{d,1}}\lambda+\text{analytic terms},
\]
so the explicit $Cu_{d,1}/\lambda$ term in
\eqref{eq:regular-EM-radial} cancels the remaining simple pole.

We next ensure that the azimuthal coefficients vanish quadratically in
the transverse radius, since the angle is undefined on the axis. On each
ordinary axis segment the scalars take constant values $\chi_A,\psi_A$.
The local potentials
\begin{equation}\label{eq:ordinary-axis-EM-patches}
A^{(A)}=A-\chi_A\dd\phi,\qquad
\widetilde A^{(A)}=\widetilde A-\psi_A\dd\phi
\end{equation}
have angular coefficients of order $\rho^2$. They extend smoothly to
that segment because, in local Cartesian coordinates
$(\xi,\eta)=(\rho\cos\phi,\rho\sin\phi)$,
\[
\rho^2\dd\phi=\xi\dd\eta-\eta\dd\xi.
\]

Near a horizon pole, apply the same construction in the regular horizon
chart:
\begin{equation}\label{eq:horizon-polar-EM-patches}
\begin{aligned}
\widehat A_{\rm in/out}^{(p)}
&=\widehat A_{\rm in/out}-\chi_p\dd\varphi_{\rm in/out},\\
\widehat{\widetilde A}_{\rm in/out}^{(p)}
&=\widehat{\widetilde A}_{\rm in/out}-\psi_p\dd\varphi_{\rm in/out},
\end{aligned}
\end{equation}
where $\chi_p,\psi_p$ are the constant scalar values on the axis segment
ending at that pole. For the upper horizon, these values are
\[
(\chi_p,\psi_p)=
\begin{cases}
(0,0),&u=1,\\
(-2Q_m,-2Q_e),&u=-1.
\end{cases}
\]
The angular coefficients $\chi-\chi_p$ and $\psi-\psi_p$ are analytic
multiples of $1-u^2$. With $\varphi=\varphi_{\rm in/out}$ and
\begin{equation}\label{eq:pole-Cartesian-coordinates}
\xi=\sqrt{1-u^2}\cos\varphi,\quad
\eta=\sqrt{1-u^2}\sin\varphi,
\end{equation}
the same Cartesian identity gives smooth angular terms at the pole.
The remaining components are analytic, and the closed gauge shifts
preserve the field strengths on overlaps.

Section~\ref{sec:balance} establishes axis regularity away from the
horizon. The ingoing and outgoing charts above extend across the horizon
away from its poles. We now check regularity at the poles, where the
horizon meets the axis. Put $s_u=1-u^2$ and define
\[
\mathsf b=\frac{\lambda^2HG}{n^2},\qquad
\mathsf c=\frac{X}{Hs_u}=\frac{\lambda^2}{Hn^2}.
\]
These coefficients are analytic for small signed $\lambda$, including
the poles.
On a section of constant $v$ (or $w$) and $\lambda$, the angular metric is
\[
q_\lambda=\frac{\mathsf b}{s_u}\dd u^2
+\mathsf c\,s_u\dd\varphi^2.
\]
In the Cartesian coordinates \eqref{eq:pole-Cartesian-coordinates},
\[
\dd\xi^2+\dd\eta^2=\frac{u^2}{s_u}\dd u^2+s_u\dd\varphi^2,
\qquad \xi\dd\xi+\eta\dd\eta=-u\dd u.
\]
Since $\mathsf b-\mathsf c=s_u a(\lambda,u)$ with analytic $a$,
the angular metric becomes
\[
q_\lambda=\mathsf c(\dd\xi^2+\dd\eta^2)
+\frac{a+\mathsf c}{u^2}
(\xi\dd\xi+\eta\dd\eta)^2.
\]
On the horizon, near either pole, $u=\pm\sqrt{1-\xi^2-\eta^2}$,
so both coefficients are analytic. At $\xi=\eta=0$, the metric reduces
to $C(\dd\xi^2+\dd\eta^2)$, establishing smoothness of the horizon
angular metric.

The horizon expansions and axis balance extend this Cartesian regularity
to a neighborhood with small positive and negative $\lambda$, including
the mixed metric components. By Eq.~\eqref{eq:horizon-nondegeneracy},
the metric determinant in the Cartesian charts tends to $-C^2\ne0$ at
either pole. Together with the polar gauge patches, this completes the
analytic Einstein--Maxwell extension across the full horizon.
\bibliography{references}
\end{document}